\documentclass{lmcs}
\usepackage{enumerate}
\usepackage{url}

\newcommand {\defin}[1]{\emph{#1}}

\newenvironment{deflist}[1]%
{\begin{list}{}%
{\settowidth{\labelwidth}{#1}%
\setlength{\leftmargin}{\labelwidth}%
\addtolength{\leftmargin}{\labelsep}%
}}%
{\end{list}}

\newenvironment{deflisteng}[1]%
{\begin{list}{}%
{\settowidth{\labelwidth}{#1}%
\setlength{\leftmargin}{\labelwidth}%
\addtolength{\leftmargin}{\labelsep}%
\setlength{\topsep}{0pt}%
\setlength{\itemsep}{0pt}%
}}%
{\end{list}}%

\newcommand {\DL}{\begin{deflist}}
\newcommand {\EDL}{\end{deflist}}
\newcommand {\DLE}{\begin{deflisteng}}
\newcommand {\EDLE}{\end{deflisteng}}

	\newcommand{\ie}{i.e.\ }
	\newcommand{\eg}{e.g.\ }
	\newcommand{\Eg}{E.g.\ }
	\newcommand{\wrt}{w.r.t.\ }
	
	\newcommand{\resp}{resp.\ }
	\newcommand{\esp}{esp.\ }

\newcommand {\typ}{\colon} 
\newcommand {\fun}{\mathbin{\to}}    
\newcommand {\pro}{\times}    

\newcommand {\la}{\lambda}

\newcommand {\ta}{\tau}
\newcommand {\io}{\iota}
\newcommand {\ro}{\rho}

\newcommand {\program}[1]{\mbox{\textup{\textsf{#1}}}}

\newcommand {\0}{\program{0}}        
\newcommand {\1}{\program{1}}        
        
\newlength{\botw}
\newcommand {\pif}{\mathop{\program{if}}}

\newcommand {\pthen}{\mathbin{\program{then}}}
\newcommand {\pelse}{\mathbin{\program{else}}}

\newcommand {\re}{\to}        

\newcommand {\sem}[1]{[\![ #1 ]\!]}  

\newcommand {\set}[2]{\{\, #1 \mid #2 \,\}}  

\newcommand {\sub}{\subseteq}

\newcommand{\co}{\mathord{\rightarrow}}
\newcommand{\Aca}{$A\co a$}
\newcommand{\Acam}{A\co a}
\newcommand{\ide}{\equiv}

\newcommand{\true}{\program{true}}
\newcommand{\false}{\program{false}}
\newcommand{\iffe}{\Leftrightarrow}
\newcommand{\rthen}{\vdash}

\newcommand{\suc}{\program{suc}}
\newcommand{\an}{\program{\,ann}}
\newcommand{\ann}{\program{\,ann}}
\newcommand{\ze}[1]{#1 \0}
\newcommand{\zee}[1]{(#1 \0)}
\newcommand{\su}[1]{(#1 \suc)}
\newcommand{\inv}{^-}
\newcommand{\limit}{\operatorname{limit}}
\newcommand{\size}{\operatorname{size}}
\newcommand{\Coat}{\operatorname{Coat}}
\newcommand{\inte}{\operatorname{int}}
\newcommand{\Gr}{\operatorname{Gr}}
\newcommand{\Con}{\operatorname{Con}}
\newcommand{\Conc}{\operatorname{Conc}}
\newcommand{\algor}{\operatorname{algo}}
\newcommand{\brac}[1]{\langle #1 \rangle}
\newcommand{\val}[1]{|#1|}
\newcommand{\cex}{\mathord{\uparrow}}
\newcommand{\csub}{\mathord{\swarrow}}
\newcommand{\din}{\mathord{\downarrow}}

\newcommand{\conu}{constructed number}
\newcommand{\conus}{constructed numbers}
\newcommand{\xyy}{(x+y^0)-y^1}
\newcommand{\aun}{a_1,\dots,a_n}
\newcommand{\aunio}{a_1:\io,\dots,a_n:\io}

\newcommand{\sme}{\simeq}
\newcommand{\smec}[1]{[#1]_{\sme}}

\newcommand{\ree}{\mathrel{\underline{\rightarrow}}}
\newcommand{\wree}{\mathrel{\underline{\mapsto}}}
\newcommand{\dree}{\mathrel{\underline{\mapsto}}}
\newcommand{\subal}{\prec}

\begin{document}
\title[Intensional Constructed Numbers]
{Intensional Constructed Numbers:\\ Towards Formalizing the Notion of Algorithm}
\author{Fritz M\"uller}
\address{Saarland University, Department of Computer Science, Campus E1.3, 66123 Saarbr\"ucken, Germany,
 \url{http://www.rw.cdl.uni-saarland.de/~mueller}}
\email{\texttt{($\la$x.muellerxcdl.uni-saarland.de)@} }

\keywords{notion of algorithm, intension, reversible program, copy mechanism, locativity}
\subjclass{F.4.1}
\begin{abstract}
This work is meant to be a step towards the formal definition of the notion
of algorithm,
in the sense of an equivalence class of programs working ``in a similar way''.
But instead of defining equivalence transformations directly on programs,
we look at the computation for each particular argument and give it a structure.
This leads to the notion of ``constructed number'':
the result of the computation is a constructed number whose constructors (0, successor)
carry a history condition (or trace) of their computation.
There are equivalence relations on these conditions and on constructed numbers.
Two programs are equivalent if they produce equivalent constructed numbers for each
argument.
\end{abstract}

\maketitle
\renewcommand{\sfdefault}{cmss}

\section{Introduction: is there a definition of ``Algorithm''?}
\emph{\textbf{Computation} still remains a fundamental mystery of the human mind}.
What fundamental things do we already know about Computation, roughly,
from the standpoint of a ``semantical'' computer scientist?

We know how to program and how to compute.
This is Turing's analysis \cite{Turing},
leading to the notion of computable function and recursion theory.
We know how to give mathematical meaning to programs,
this is Scott's denotational semantics \cite{Scott}.
We know how to formally prove qualitative properties of programs,
this is Hoare's program logic \cite{Hoare}.
There is complexity theory exploring two quantitative properties of programs \cite{Hartmanis/Stearns}.
Complexity theory has some unsolved fundamental problems 
for which we do not even know the theoretical means to attack them \cite{Gasarch:1,Gasarch:2}.
There is a deep division between the qualitative and the quantitative theories
of Computation. 
 But there are already bridges between them,
mainly within the topic of implicit complexity,
where complexity classes of higher type functions are explored,
logical characterizations of complexity classes are given,
or complexity classes are characterized by syntactic restrictions
of the computation mechanism by various means.
All this remains mainly on the syntactic level,
there still is no use of denotational semantics or program logic in complexity theory.

Is there missing something else, perhaps something more obvious and simpler,
whose solution could lead to new theoretical means and insights?

Computer scientists always talk about \emph{algorithms}.\\
\textbf{So what is an Algorithm?}\\
\textbf{There still is no general formal definition of this notion}. 
\begin{itemize}
\item Please note that an algorithm is \textbf{not a program}, and not a Turing machine.
And a program does not become an algorithm when it is written in natural language.
\item Please note that an algorithm is \textbf{not a computable function}. 
\item Intuitively, an algorithm is some \textbf{equivalence class of ``similar'' programs}
for the same computable function.
The problem is to define the corresponding equivalence relation on programs.
\end{itemize}
If you think that ``insertion sort'' and ``bubble sort'' are different algorithms,
then you should be able to prove this.
(I have chosen these two algorithms with the same time complexity,
so that their difference cannot be justified by different complexity.)

But generally, there is almost no awareness of the problem.
There are only few people, notably Yiannis Moschovakis \cite{Moschovakis:what}
\cite{Moschovakis:found}, Noson Yanofsky \cite{Yanofsky}, and Bergstra/Middelburg
\cite{Bergstra/Middelburg},
who have insisted on posing the problem and giving partial answers.
Besides these works, the talk about algorithms has just produced some kind of ``bureaucracy''
where certain important algorithms (programs) are sorted out
and named.
By this talk we all have subconscious intuitions, a kind of ``feeling'',
when two programs should be regarded as equivalent.

What should/could/might be expected from a formalization of the notion
of algorithm and its theory?
\begin{itemize}
\item First, the definition should be more scientific than bureaucratic.
\item There may be different notions of algorithm, for different aspects or 
different purposes, or even a whole spectrum of notions.
The topic seems to have a great arbitrariness, at least in the beginning.
\item The notion of algorithm should be independent of the language in which
the programs are written,
\ie the notion should be unique for the data domain at hand.
But for the beginning it would be enough to give a definition for
a particular programming language.
\item There may be counter-intuitive results:
It might come out that ``insertion sort'' and ``bubble sort'' are the same algorithm.
Or there might be two programs with different time compexity that are
equivalent as algorithms.
(Counter-intuitive results are no exceptions in science, you will easily find examples.)
\item And even worse: 
It might turn out that the notion of algorithm does not directly refer
to the ``normal'' (Turing machine, imperative, functional) programs,
but to programs of a different kind,
think of reversible or quantum programs.
(In fact, our tentative solution is about a new kind of programs
that have much of the flavour of reversible ones.)
\end{itemize}

I do not bother much whether the definition that may come out matches
the intuitions for algorithms.
Who cares about a definition that nobody seems to need?
For me the importance of the question is as an indicator that
something fundamental is missing, has been overlooked.
So I think the question  is just a good starting point to dig deeper,
it is a guidance for a new analysis \emph{per se} of Computation.

This introduction is continued in the next section with the basic idea
of our tentative solution of the algorithm question,
the constructed numbers.
Please note that the discovery of the constructed numbers
(and their generalizations)
is the main achievement of this paper.
We are still far from the full solution of the formalization of ``algorithm''.
This paper is a preliminary version.
At the end of the next section there is an outline of the paper.

\section{The basic idea of constructed numbers}\label{s:emp2}

We begin our analysis with another question different from the algorithm question,
it could be something like:
\begin{center}
\emph{Do we fully know what is happening when we compute?}
\end{center}
(Please note that here humans are supposed to compute,
sorry for being so old-fashioned.)\\
I admit that this is a silly question,
as it has been answered already in 1936.
But let us try a brief fresh analysis of it:

In a Computation, there are data elements of some ontology,
here numbers typically described by Peano arithmetic.
We think mainly of functional Computation.
There are different qualities of Computation steps.
There are constructions ($1+1$), there are destructions ($1-1$),
and there are conditions/decisions/observations ($\pif x=0 \pthen \dots \pelse \dots$).
Recursion and the execution of conditions accounts for the dynamics of the Computation process.

The reality that we see at some point (of time and of place)
in the execution of a program are the values of the variables at this point;
these values are taken from the data elements.
This reality corresponds to the unconscious moves of the mechanic computer.
I call it the objective reality.
It is the natural, primary reality.
Both systems of making sense of programs,
Scott's denotational semantics and Hoare's program logic,
take this reality as their basis.

But in the world there are always different realities.
First, often the values of the program are interpreted in some
material reality external to the program.
But also when we work just inside
a formal or programming language,
we may impose any interpretation we like as long as it is in accordance
with the structure of the language, in some sense.
Computer scientists know this already:
they may interpret their programs in different ways,
with values of some other kind,
\eg with more abstract values in abstract interpretation \cite{Cousot}.

\smallskip
So let us be insolent and criticize the objective reality of Computation,
and ask if there is a fuller reality that encompasses the primary one;
and the \emph{critique} goes like this:
\begin{quote}
The value of a variable at a program point should not just be a plain number,
but a \emph{constructed number},
\ie a number that carries the history of its construction or deduction.
The value of the variable should become dynamic, as Computation is.
\end{quote}

Let us develop this idea in a first approximation.
\Eg there may be a number $a$ that is a ``$1$'' that has been constructed,
caused by the fact that the condition $x=0$ was checked.
Let us symbolize it as $a=((x=0)\co 1)$.
And there is a similar number $b$ that is a ``$1$'' after the check of $y=0$,
where the (program) variable $y$ is different from $x$,
$b=((y=0)\co 1)$.%

Now the forgetful computer, already unconscious of its former constructions
of $a$ and $b$ by conditions,
may compute $a-b=0$.
But is this ``correct''?
It is correct in the \emph{extensional} sense,
the \emph{extension} of a constructed number $a$ being the corresponding plain number,
expressed as $\val{a}$.
It is $\val{a}=\val{b}=1$ and $\val{a}-\val{b}=0$.

It is \emph{not correct} in the \emph{intensional} sense,
the \emph{intension} of a constructed number being \dots
the constructed number expression itself modulo some equivalences.
It is not $a=b$, and not $a-b=0$,
because $a$ and $b$ are (intensionally) different,
and the difference of two different numbers cannot be $0$.

So the (human!) computer has to go back to school again (sorry for that),
and learn how to compute intensionally.
What should she do with the expression $a-b$?
Simply let it stand as it is for later use,
it cannot be reduced intensionally itself.
It is an expression that has to be carried through the Computation
in suspended unreduced form.
Later the constructed number $b$ may come again and we may have
to compute $(a-b)+b$.
Then we get $(a-b)+b=a+(b-b)=a$,
and this is intensionally correct.
The general arithmetic laws stay valid for intensions.

So the subtraction of identical numbers,
\ie with the same history, $b-b=0$, annihilates,
while the subtraction of non-identical numbers, $a-b$,
with different history, does not.
This is one of the main points of our definition of algorithm:
Annihilation in all cases would lead to the final result,
so to the computed function of the program.
Annihilation only in cases of identity leads to a finer distinction
of programs, to algorithms.

\smallskip
In our first approximation we have seen constructs \Aca, where $A$
is a \textbf{propositional formula} $(x=0)$
and $a$ is a number.
We keep this variant of our new constructions in mind for the \emph{next section}.
Our constructed numbers will in fact be another variant,
where $A$ is a \textbf{history} (or trace or certificate) of Computation.
We want to get rid of the variable $x$ in $(x=0)\co 1$.
And we want to write our numbers in constructor form,
with the constructors $\0$ and $\suc$ (and another constructor),
and associate the conditions to the constructors.
The history-conditions are, roughly, formal products of atomic conditions.
The value $\0$ of the variable $x$ will itself be conditioned with its history $A$.
So the value of $x$ is of the form $\ze{A}$,
where $A$ is the condition, the history, of this occurrence of $\0$,
and simply written before the constructor $\0$.
The result of the Computation, the conditioned $\1$,
will now be written $\su{A^1}\zee{A^0}$,
where we make two copies $A^1$ and $A^0$ of the condition $A$,
and write them before the constructors $\suc$ and $\0$.
The variable $x$ no more appears in this expression.

So the conditions of our constructed numbers are not propositional formulas,
but histories (or traces or certificates) of Computation.
They are described not by logic,
but by a new kind of algebra.

This is in a very small nutshell the idea of the arithmetic CN of
constructed numbers,
which will be given in Sections \ref{s:cnexp} and \ref{s:cn}
with more features than described here.
The expressions  of CN
arose as traces of Computation.
The main insight was to see that these are numbers in their own right.

\smallskip
What have \conus\ to do with algorithms?\\
The system CN of constructed numbers is the main achievement of this paper
and it was motivated by the algorithm question.
The common approach to the algorithm question is to define
equivalence transformations directly on programs,
\eg \cite{Yanofsky}.
We take another approach:
we look at the Computation for each particular argument
and give it a structure.
The result of this Computation is a \conu\ and the conditions of its constructors
carry the history of their computations.
There are equivalence relations on conditions and on constructed numbers.
Two programs are equivalent if they produce equivalent constructed numbers
for each argument, see Definition \ref{d:algo}.
But there is a difficulty:
with the introduction of \conus\ the programming has changed its
character in CN,
it is programming with numbers \emph{and with conditions}.
It looks more like reversible computing 
in that it also keeps traces.
There should be a notion of correct correspondence between CN and normal programs,
but we must leave this question open for the moment.

\smallskip
\emph{What has been achieved, in general terms?}\\
We have made a whole Computation an object
(in the form of the conditioned final result),
and given it some specific structure.
The most primitive form of doing this is to take just the sequence of computation
or reduction steps and to count these steps, to get  just a number.
We can read out more information from our solution,
\eg we can see which input observations were needed for the generation
of a particular constructor of the result, and much more.
I only know of two other approaches to make a Computation a structured object:
the study of the equivalence of reductions of term rewriting systems \cite{Oostrom/Vrijer},
and Jean-Yves Girard's research programme of Geometry of Interaction
\cite{Girard:tgoi}\cite{Girard:goii}\cite{Haghverdi/Scott},
which is primarily on the normalization of logic proofs.
There is a certain arbitrariness and freedom to choose what we want to see in the structure
of a Computation.

\smallskip
\emph{What will be the use of our system CN of constructed numbers}, besides
being a step towards formalizing ``algorithm''?
Can we program better?
First, to keep things simple,
our tentative solution is only about numbers in tally form.
This is not enough to do normal complexity theory.
Lists have to be coded.
But our principles can surely be extended to a system with list
or tree data types.
Even then,
programming in CN is a different kind of programming
and more complicated,
so that it is not meant to compete with conventional programming on
its own ground.
But the applicability might change with the advent of reversible and quantum computing.
CN has the explicit keeping of computation traces in common with reversible programs,
but a CN program need not be reversible.
Another application could be in analyses of programs connected to complexity.

\subsection*{Outline of the paper:}
\begin{enumerate}[1.]
\setcounter{enumi}{2}
\item Subjective objects: ideas and philosophy:\\
Here we take up the idea of our ``first approximation''
(with propositional formula $A$ in \Aca) of Section \ref{s:emp2}.
Our ``subjective objects'' are general constructs $A\co a$,
where the condition $A$ is a history (as for our constructed numbers)
or a propositional formula.
We explain the meaning of them 
and speculate about a ``decomposition'' of set theory.
This section is a speculative digression from the algorithm topic and is more on logic.
It can be skipped by those interested only in algorithms.
\item
System CN (constructed numbers): condition algebra and explanation with examples:\\
We explain the programming in CN with the examples of addition and subtraction.
In Subsection \ref{s:cond}
``Finite size limitation of Computation, the condition algebra and the bracket mechanism''
we also explain a copy mechanism that was inspired by
Jean-Yves Girard's locativity \cite{Girard:locus,Girard:locat}.
\item
System CN (\conus): the arithmetic:\\
Here the system CN is completed by giving the rules on numbers.
\item
Basic notions of the theory of system CN:\\
We give the notions of CN-algorithm and direct computation.
\item
Outlook
\end{enumerate}

\section{Subjective objects: ideas and philosophy}

In the last section we saw constructed numbers $A\co a$ (read ``A condition a''),
where $A$ is a fact about the (history) of Computation,
and $a$ is a natural or constructed number.
The meaning is that $A$ is a condition for the existence of the individual $a$ constructed.
Already before giving the formal system CN,
let us \emph{speculate} here about the deeper meaning and possible generalizations of these constructs.
Please note that we have only realized the system CN in this paper.
This section can be skipped by those interested only in algorithms. It is more on logic.

\smallskip
The constructs \Aca\ come in two guises,
with history conditions $A$ or with propositional conditions $A$.
I like to call all these objects \Aca\ generally ``subjective objects''.

Why ``subjective objects''?\\
The polarity between subject and object is primordial for the human mind.
We always think about objects,
and we do this always by subjective means.
There is an interplay between the two realms,
nothing can be seen in one-sided isolation.
So pure objects are only an idealization.
In mathematics, 
the logical formulas and the processes of proof, deduction
and computation at first sight belong to the realm of the subjective.%
\footnote{ It should be clear that the adjectives \emph{objective} \resp 
\emph{subjective} have just the meaning
``belonging to the realm of the object \resp the subject'',
not the popular meaning ``being generally valid or formal'' \resp 
``being valid only for one person or informal''.}

Therefore the name ``subjective object'' for \Aca,
as we form a new object from the object $a$
by adding a (subjective) logical formula $A$ \resp 
a (subjective) history of a computation.

So there is no fixed border between the objective and the subjective,
the border can be shifted.
Subjective processes can be construed as objects,
they can be \emph{reified} (\emph{verdinglicht}),
as is done \eg in proof theory.
Think also of G\"odel's coding of propositions as numbers.

\smallskip
At last we come to the decisive question:\\
What does the subjective object \Aca\ ``mean''?\\
We must distinguish the two cases:

\smallskip
\noindent
\textbf{(1)} $A$ is a history of a computation:\\
\Aca\ does \emph{not} stand for a set-theoretic object,
just as a program does \emph{not} stand for a set-theoretic object.\\
For numbers the meaning will be given by the rules of system CN,
which describe how we can compute with constructed numbers.

\smallskip
\noindent
\textbf{(2)} $A$ is a propositional formula (for the rest of this section):\\
Also here \Aca\ does \emph{not} stand for a set-theoretical object.
At best a set-theoretical interpretation of the system can be given,
\esp when we are in a simple system where $a$ is a number.
This interpretation assigns for each variable-environment $\ro$
(a function from the variables to normal numbers)
a set of normal numbers
\[ \sem{\Acam}\ro = \pif \sem{A}\ro \pthen \sem{a}\ro \pelse \emptyset, \]
roughly.

The general informal meaning of \Aca\ is hard to describe.
The problem is that we build a new kind of object that falls out
of the basic universe of objects on which all is grounded (normal numbers, sets).
So it does not suffice to use the language of these basic objects.
We must circumscribe the meaning in a kind of contradictory way;
contradictory not in the formal logical sense,
but in the sense of ``contradictory in the notion'',
just as when we say ``light is a wave \emph{and} a particle''.
I have to offer the following two meanings'' of \Aca:\\
\textbf{(a)}
The object $a$ which comes to existence when $A$ is fulfilled, or\\
\textbf{(b)}
The concept of (an object $a$ under the fulfillment of $A$),
but this concept construed as an object.\\
So \Aca\ is in some sense both an object and a concept.
Let us call these objects of variant~(2) ``concept objects''.
I admit that this is very strange,
but I have some (preliminary) rules that should express precisely the meaning of \Aca,
the most important ones are:\\
$(\true\co a)\ide a$\\
$A\co (B\co a) \ide (A\land B)\co a$\\
$(A \iffe B)\land a\ide b \rthen (\Acam)\ide (B\co b)$\\
$B[x:= A\co a] \rthen (A\co a)\ide (B[x:=a]\land A)\co a$\\
from this follows: $B\rthen (\Acam)\ide (B\land A)\co a$\\
from this follows: $B\rthen a\ide(B\co a)$\\
If we are in a system with numbers,
there are also rules to compute, \eg:\\
$(\Acam)+ (A\co b) \ide A\co (a+b)$

\smallskip
Of course, subjective objects are not new.
But the old forms always had a plain meaning in the basic universe of objects,
they did not transcend it.
There are the usual $\epsilon$ or $\iota$ description operators.
Please note that our construction is \emph{not} of this kind.
\Aca\ is \emph{not} the object $a$ that fulfills $A$.
This latter object does not exist when $A$ is not fulfilled.
\Aca\ always exists, regardless whether $A$ is fulfilled or not.
\Eg $(\false\co 0)$ is a very honourable number,
it is not $0\ide(\false\co 0)$,
but it is $(\false\co 0) \ide (\false\co 0)$.

\smallskip
The expressions of set comprehension in set theory are also examples
for subjective objects:
$\set{x}{A}$ should be the set of all $x$ for which $A$ is fulfilled,
if this set exists.
Georg Cantor's informal definition of set was this:
\begin{quote}
A set is a multitude of things that can be thought as a unity.
\end{quote}
Guided by this definition we can make a decomposition of the object $\set{x}{A}$
into three steps:
\begin{enumerate}[(1)]
\item
We start with our subjective object $A\co x$.
This is just the object that denotes the concept of an $x$ fulfilling $A$.
\item
We make of this object the multitude of all such $x$ by the expression
$\delta x.(A\co x)$,
where $\delta$ is a ``data choice operator''
binding the $x$ in \Aca.
(This is not yet a unity.)
\item
We have convinced ourselves that this multitude can be thought as a unity.
This does not mean that it \emph{is} already a unity.
We must \emph{make} it a unity, if we want to:
$\{ \delta x.(A\co x)\}$.
\end{enumerate}
Based on these constructions,
there (hopefully) will be a general theory expressing concept objects,
multitudes, classes and sets.
The property of being a set is definable in this system.
What about Russell's paradox?
We can form the class
\[ \delta x.((x\text{ is set and }x\notin x)\co x), \]
which is not a set.

\section{System CN (constructed numbers): condition algebra and explanation with examples}\label{s:cnexp}

We have explained the basic idea of \conus\ in Section \ref{s:emp2}.
Here we try a gentle introduction to CN guided by two example programs of addition and
subtraction,
like an introduction to a new programming language.
It should be clear that we explain new features and rules when they are needed on the way,
there will always be points left open.
The algebra of conditions is given here in full detail in subsection \ref{s:cond},
the rest of CN appears in full detail in Section \ref{s:cn}.

First, the normal numbers are built by the constructors $\0$ and $\suc$,
and the programs are non-deterministic first-order recursive reduction rules
for each defined function.
Here is a normal program for addition,
$+\typ \io \pro \io\fun \io$,
$\io$ integer type, $x,y$ number variables:
\begin{alignat*}{2}
(\suc\, x)&+ y &&\re \suc(x+y)\\
\0 &+ y &&\re y
\end{alignat*}
This was too simple,
let us adapt this definition to \conus.
There are condition expressions $A,B$ and number expressions $a,b$.
Conditions are built up from 
atomic conditions by an algebra with a formal product being associative and commutative,
and other operators.
Imagine the atomic conditions as free objects of the algebra,
we will not see them in this section.
The basic numbers are built up from the constructors
$A\0$, $A\suc$ and $(A,B \ann)$,
where $A,B$ are condition arguments of the constructors.

We have a third constructor ``$\an$'' which takes two condition arguments $A,B$
and a number argument $a$: $(A,B\an)a$.
In the programs, $(A,B\an)$ is created from the (suspended) mutual annihilation 
of $A\suc$, taken positively,
and $B\suc$, taken negatively.
Such a creation takes place \eg in a subtraction,
which we will see below.
$(A,B\ann)$ behaves extensionally as the identity function.
$(A,B\ann)a$ can be reduced to $a$ when $A,B$ are ``inverses'' of each other,
in a certain sense that is not the sense of groups.
In the other cases $(A,B\an)$ has to be carried through the Computation,
but it can react and be observed and processed in a reduction.
This means that the programs must have reduction rules also for the case of $\an$,
if they are not under-specified.
(But functions are allowed to be under-specified.)

There are six relations on conditions and numbers:
\begin{itemize}
\item
The equality $=$ on conditions and the ``smooth equality'' $\sme$ on numbers
of basic equations (congruent equivalences).
These are independent of the program.
\item
The reduction relation $\re$ on numbers caused by the program rules (a congruence).
\item
The ``equality reduction'' $\ree$ on numbers that encompasses $\sme$ and $\re$
(reflexive, transitive and congruence).
\item
The ``direct equality reduction'' $\dree$  on numbers is a restriction of $\ree$
that accounts for ``direct'' Computation.
\item
In case of consistency (of reversing the reduction rules, see below) there is the equality $=$ on numbers defined by 
$a=b$ iff $a\ree b$ and $b\ree a$.
\end{itemize}
Examples are:
\[ AB=BA, \qquad (AB)C= A(BC), \text{ for conditions} \]
\[ (A\suc)(B\suc)a \sme (B\suc)(A\suc)a, \]
exchange laws on numbers for any pairwise
combination of $\suc$- and $\ann$-constructors.\\
The conditions give the individual constructors an identity,
and our aim is to give them each a unique identity by a unique condition, roughly.
With the exchange laws the sequence of the constructors of a number can be permuted
in any order.
So we can push the ``right'' constructor to the top in order
to perform a reduction rule for a function.
Programming in CN means programming with numbers \emph{and with conditions}.

\smallskip
Here is a possible addition program for \conus\ derived from the normal program above,
the $X,Y$ are condition variables:
\begin{alignat}{2}
\label{a1}(X\suc)x &+ y &&\re (X\suc)(x+y)\\
\label{a2}(X_0,X_1\an)x &+ y &&\re (X_0,X_1\an) (x+y)\\
\label{a3}X\0 &+ (Y\suc)y &&\re (Y\suc)(X\0 + y)\\
\label{a4}X\0 &+ (Y_0,Y_1\an) y &&\re (Y_0,Y_1\an)(X\0 + y)\\
\label{a5}X\0 &+ Y\0 &&\re \brac{XY}\0
\end{alignat}
Here the constructor $\an$ is treated like $\suc$:
it walks up out of the sum unchanged.
There is the same case analysis in the first argument of $+$ as in the normal program.
But for the case $X\0$ there is a second recursion over the second argument of $+$
to bring $X\0$ down to $Y\0$.
In the end, they coalesce to $\brac{XY}\0$.
If we intensionally change a rule,
\eg $X\0 + Y\0 \re Y\0$ as the last rule,
then we still have an addition program in the extensional sense,
but with different (intensional) properties.
In this example the commutativity of addition would get lost.

\smallskip
Here is a new operator of the condition algebra:
$\brac{A}$, the bracket operator.
It is always used to enclose the composed condition of a constructor in the right rule side
(here $\brac{XY}$),
so that it has limited capabilities to react with the outside. 
Why that?

\subsection{Finite size limitation of Computation, the condition algebra and the bracket mechanism}\label{s:cond}
$\phantom{x}$\\
Since 1936 we know that Computation has a finite size limitation.
In his analysis Alan Turing explained that a computer has only finitely many states of mind,
so that her consciousness for the Computation is limited.
Accordingly, the Turing machine has finitely many states,
and the program has finite size.
The state changes, when data are observed:
state $s$ $\re$ atomic datum $d$ $\re$ state $s'$.
There are states that are fully ``conscious'' of a \emph{finite}
sequence of preceding data that have been observed;
and states that lack such knowledge,
because they have already ``digested'' the data.
In a Turing machine, the different quality of these states does not appear in the syntax, it is unstructured.

In a program, \esp a functional one, the different quality is distinguished in the syntax:
the ``conscious'' states appear after observations in an if-then-else construct.
In our programs they appear after a whole left rule side is matched,
and the gathered knowledge comprises all the constructors that have been matched.
There are pieces of this knowledge distributed on the constructors of the right side
by our condition mechanism.
We give these pieces of information/condition,
that were fully known at that moment of creation,
a special status by enclosing them in brackets $\brac{A}$;
so that the laws on conditions, like commutativity and associativity,
are only applicable inside the brackets and cannot ``cross the border''.

So we have a trace of the recursive structure of the Computation in the conditions
(of the final result).
What would happen if we had no bracket operator?
Then all the atomic conditions in the condition of an output constructor
would be mixed together in a big pot by associativity and commutativity
and can annihilate and merge,
so that in the end the condition just says which input constructors were used
for the output constructor.
The recursive structure of the Computation would get lost
and we merely have a quantitative information.

But we do not want the borders of the brackets to be strict,
we want to be allowed to shift them.
The reason is that we want to identify some programs (for the same function) as algorithms,
but of cause not all of them.
So we introduce the rule
\[ \brac{A}\brac{B}=\brac{AB}.\]
(I have also tried other rules, namely $\brac{A}B=\brac{AB}$ \resp
$\brac{A}=A$, but dismissed them.)\\
But this alone does not make it.
The condition inside a bracket $\brac{A}$ can get unlimited many factors,
and we have seen above that in a Computation this size is always limited.
Hence our system CN comes with a parameter ``$\limit$'' ($\geq 3$),
it must always be $\size(A)\leq \limit$.
The parameter $\limit$ is set once for each proof that is performed in CN.
(But in many cases it need not be set to a fixed number as $\limit\geq 3$ is sufficient.)
The rule $\brac{A}\brac{B}=\brac{AB}$ is not applicable when $\size(AB)>\limit$.
What would happen if we had no size limitation?
Things like described above.
Programs would become equivalent that cannot be justified so by local transformations.

\medskip
We now give the complete algebra of conditions.
Conditions are built up from atomic conditions (of a countably infinite set $\Coat$)
and variables by some algebraic operations.
The atomic conditions obey no other laws than those given here.

We already said that we want to distinguish each individual constructor in a number. 
This is a variant of the idea of ``locativity'' of Jean-Yves Girard \cite{Girard:locus,Girard:locat}.
For this, system CN has a built-in copy mechanism which makes out of a number term $a$
two copies $a^0,a^1$, and out of a condition term $A$ two copies $A^0, A^1$.
We explain how a term has to be.
\begin{defi}[copy exponent]\label{d:cex}
We can define the notion of \defin{position} $p$ in a term $t$ (a condition or number term)
in the usual way as a word over a small alphabet.
$t/p$ is the subterm of $t$ at position $p$.
We define the \defin{(copy) exponent} of position $p$ in $t$, $t\cex p$,
as the sequence of exponents $0,1$ 
that we see on the way from $p$ walking up to the root of $t$.
\Eg let $t=((X^{0-}Y\suc)^1 y^{10} + z)^0$.
Then for $p$ the position of the occurrence of $X$ in $t$ we get $t\cex p = 010$.
For $q$ the position of the occurrence of $y$ in $t$ we get $t\cex q = 100$.
\end{defi}
\begin{defi}[unique copy exponent]\label{d:ucex}
Let $t$ be a condition or number term.
$t$ has \defin{unique (copy) exponents} if the following is fulfilled:
Let $s$ be an atomic condition, a condition variable $X$,
or a number variable $x$.
Let $p\neq p'$ be positions in $t$ and $t/p=t/p'=s$.
Then $t\cex p$, $t\cex p'$ are not comparable,
\ie neither $t\cex p\leq t\cex p'$ nor $t\cex p' \leq t\cex p$.
(Here for words $v,w$ it is $v\leq w$ iff there is a word $v'$ with $vv'=w$.)
\end{defi}
This means that two different occurrences of $s$ are distinguished by their incomparable
copy exponents.
If the term $t$ has unique exponents,
then also every subterm of it.

\medskip
The \textbf{conditions} $A,B,C,D$ are:\\
condition variables $X,Y,Z$, atomic conditions $A\in \Coat$,\\
$AB$ (product), $I$ (neutral element), $A^-$ (a kind of inverse, but not in group sense!),\\
$A^0$, $A^1$, $\brac{A}$.

The size function is defined on conditions in their purely syntactic form:
\begin{align*}
\size(I) &= 0  &  \size(AB) &= \size(A)+\size(B) \\
\size(X) &= 1, \text{ for $X$ variable}  &  \size(A^-) &= \size(A) \\ 
\size(A) &= 1, \text{ for $A$ atomic}  &  \size(A^0) &= \size(A) \\ 
\size(\brac{A}) &= 1,  &  \size(A^1) &= \size(A) 
\end{align*}
Please note that the ``condition placeholder'' $A$  and the condition variable $X$ 
of the language have different character.
$A$ is used in the laws of the algebra,
whereas $X$ is used in the reduction rules for the functions.
$A$ can be replaced by any condition,
whereas $X$ can be replaced only by conditions $B$ with $\size(B)=1$,
so that $\size(X)=1$ is valid even after replacement.

Every condition term $A$ must be \emph{limited}, \ie for all subterms $B$ of $A$ it is $\size(B)\leq \limit$.\\
Every condition term must have unique copy exponents.
For an equation to be valid,
both sides must have them.
If we make a replacement in a (condition) term according to a valid equation,
then the term stays with unique exponents.
(Please note that for these restrictions the product $AB$ is a partial operation.)

\medskip
The \textbf{equality on conditions}:
\begin{align*}
(AB)C &= A(BC)  &  (AB)^0 &= A^0B^0 \\
AB &= BA  &  (AB)^1 &= A^1 B^1 \\
AI &= A  &  \brac{I} &=I \\
A^{--} &= A  &  \brac{A}\brac{B} &= \brac{AB} \\
(AB)^- &= A^- B^-  \\
\intertext{The following equations will later have a special status
because of their asymmetric character:}
A^0A^{1-} &= I  &  A^0A^1 &= A \\
[ A^1A^{0-} &= I \text{ this can be deduced, see below} ]
\end{align*}
There are rules that close $=$ to be a congruent equivalence.\\
Perhaps we should also add
$\brac{A}^-=\brac{A^-}$,
$\brac{A}^0=\brac{A^0}$,
$\brac{A}^1=\brac{A^1}$.

For technical reasons, there are the following rules for numbers:\\
For a condition $A$ of a constructor it must always be $\size(A)=1$.
\begin{align*}
A\0 &\sme \brac{A}\0  &  (A,B\ann)a &\sme (\brac{A},B\ann)a \\
(A\suc)a &\sme (\brac{A}\suc)a   &  (A,B\ann)a &\sme (A,\brac{B}\ann)a
\end{align*}
\begin{prop}[due to Reinhold Heckmann]\label{p:heck} $\phantom{x}$\\
In the condition algebra we have:\\
(1) $I=I^-=I^0=I^1$ \\
(2) $A^1A^{0-} = I$
\end{prop}
\proof
(1) $I^-$ also is a neutral element:
$AI^-=A^{--}I^- = (A^-I)^- = A^{--} = A$.\\
$I=II^- = I^-$.\\
$I^0 = I^0I = I^0(I^0I^{1-}) = (II)^0I^{1-} = I^0I^{1-} = I$.\\
$I^1 = I$ analogous.\\
(2) $A^1A^{0-} = A^{1--}A^{0-} = (A^{1-}A^0)^- = I^- =I$.
\qed
\begin{prop}
(1) If $A\neq I$, then $A$ contains a variable or an atomic condition.\\
(2) $AA^- = I$ iff $A=I$.
\end{prop}
\proof
(1) by induction on the term $A$.\\
(2) Let $A\neq I$.
Then $A$ must contain a variable or an atomic condition, let us name this $v$.
Let $p$ be the position of some occurrence of $v$ in $A$.
Then it is $(AA^-)/(1.p) = (AA^-)/(2.1.p) = v$,
but $(AA^-)\cex (1.p) = (AA^-)\cex (2.1.p)$.
(Here $1$ means the left, $2$ the right subterm.)
So $AA^-$ has not unique exponents.
\qed

\begin{rem} (due to Reinhold Heckmann)\\
If we do not impose the restriction of unique copy exponents,
then we get the contradiction (to copies) $A^0=A^1$ for every $A$.\\
For every $A$:
$ AA^- = (A^0A^1)(A^0A^1)^- = A^0A^1A^{0-}A^{1-} =
(A^0A^{1-})(A^1A^{0-}) = II = I$\\
Then
$ A^0 = A^0I = A^0(A^1A^{0-}) =
(A^0A^{0-})A^1 = IA^1 = A^1 $

If we keep the restriction of unique exponents, 
but add the equations 
$A^{-0}=A^{0-}$ and $A^{-1}=A^{1-}$,
then we get other contradictions like
$A^{01} = A^{10}$ and $A^{00}=A^{11}$.
\end{rem}

\begin{rem}
To prove consistency of the condition algebra,
\ie absence of such contradictions,
we can make a model of normal form representations.
To keep things simple,
we leave out the bracket operator for the moment.
We give a sketch of the proof.\\
We define \defin{elementary conditions} $u$ as conditions $X^e$ or $A^e$,
where $A$ is atomic,
and exponent $e$ is a finite word over $0,1,-$. (These are not the copy exponents!)\\
A \defin{set condition} $S$ is a finite set of such elementary conditions,
which obeys the (analogous) property of having unique \emph{copy} exponents.\\
There are four reduction rules on set conditions:\\
(1) replace $--$ in an exponent by the empty word,\\
(2) replace a subset $\{u^0,u^1\}$ by $\{u\}$,\\
(3) replace a subset $\{u^0,u^{1-}\}$ by $\emptyset$,\\
(4) replace a subset $\{u^1,u^{0-}\}$ by $\emptyset$.\\
We get the normal form $nf(S)$ of $S$ by:\\
(a) reducing by rule (1) until it can no more be applied, then\\
(b) reducing by rules (2-4) to normal form.\\
The process (b) is confluent,
because there is no ``overlap'' between the rules (2-4),
because of unique copy exponents.
As it is also terminating, the normal form is unique.\\
We give an interpretation $\inte$ of the condition algebra:\\
$\inte(X)=\{X\}$, $\inte(A)=\{A\}$ for $A$ atomic,
$\inte(AB)=nf(\inte(A)\cup \inte(B))$,
$\inte(I)=\emptyset$,
$\inte(A^-)= nf((\inte(A))^-)$,
$\inte(A^0)=(\inte(A))^0$,
$\inte(A^1)=(\inte(A))^1$.
(In the last cases the exponent works on all elements of $\inte(A)$.)\\
This interpretation fulfills the equations.\\
For every condition $A$ it is $\inte(A^d) \neq \inte(A^{d'})$ for $d\neq d'$
words
of $0,1$ as exponents.
\end{rem}
\textbf{End of subsection \ref{s:cond}}

\bigskip
A problem for equality arises with the reduction rules $\re$,
which may be non-deterministic,
in both extensional and intensional sense,
so $\re$ cannot be taken in reverse direction as part of equality.
(The extensional non-determinism can be forbidden, if not wanted,
but the intensional non-determinism seems to be useful in many cases.)
We must ensure that the reverse reductions do not cause contradictions,
\ie that there are no terms $a,b$ without function symbols for which we can deduce
$a=b$ though not $a\sme b$.
Only then can we establish $=$ as the equality encompassing $\sme$ and $\re$.
This can often be proved by confluence.

In the case of our addition program,
we use the confluence Theorem 3.3 of \cite{Huet},
with the complete set of laws in Section \ref{s:cn}.
Essentially, we must check termination of the composition of $\re$
with one step of $\sme$;
and the convergence of all critical pairs that are caused by an overlap
of two rules of $\re$ (there are none),
or by an overlap of a rule of $\re$ with a rule of $\sme$ (there are some).

Having established the equality $=$ for our addition program,
we can prove the equalities
$x+y=y+x$ and $(x+y)+z = x+(y+z)$
for all \conus\ $x,y,z$,
by induction on \conus\ in their constructor form.
The constructor form is the form built from $A\0$, $A\suc$, $(A,B\an)$.
These are the basic objects that exist.

For the proof of $x+y=y+x$:\\
As the constructors $\suc$ and $\an$ behave in the same way,
these cases are analogous.
In the outer induction over $x$ there are two inner inductions 
to prove $(X\suc)(y+x')=y+(X\suc)x'$
and $X\0 + y=y+X\0$.
All the exchange laws for constructors are used 
and the commutativity of condition product $XY=YX$.

For the proof of $(x+y)+z= x+(y+z)$:\\
In the outer induction on $x$ there is an inner induction to prove
$(X\0+y)+z= X\0+(y+z)$,
and in this induction there is an inner induction to prove
$(X\0+Y\0)+z = X\0 + (Y\0+z)$.
Associativity on conditions is used.

\bigskip
Now to natural number subtraction $-\typ \io\pro\io\fun\io$.
A normal program is this:
\begin{alignat*}{2}
(\suc\, x) &- (\suc \, y) &&\re x-y\\
x &- \0 &&\re x\\
[\0 &- (\suc \, y) &&\re \0]
\end{alignat*}
Here is a possible subtraction program in CN:
\begin{alignat}{2}
\label{s1} (X\suc)x &- (Y\suc) y &&\re (X,Y\ann)(x-y)\\
\label{s2} x &- (Y_0,Y_1\ann)y &&\re (Y_1,Y_0 \ann)(x-y)\\
\label{s3} (X\suc)x &- Y\0 &&\re (X\suc)(x-Y\0)\\
\label{s4} X\0 &- Y\0 &&\re \brac{XY\inv}\0\\
\label{s5} (X_0,X_1\ann)x &- y &&\re (X_0,X_1\ann)(x-y)\\
\label{s6} [X\0 &- (Y\suc)y &&\re X\0]
\end{alignat}
The rule \eqref{s1} is the rule where a constructor $\ann$ is created from
the subtraction of two $\suc$s.
Rule \eqref{s2} forms the ``inverse'' of $(Y_0,Y_1\ann)$ by reversing the order
of the two conditions.
In rule \eqref{s4} the inverse of $Y$ is employed.

We have set the rule \eqref{s6} in brackets, as it destroys confluence of the program.
The program without this rule should be confluent,
but I do not yet know how to prove it.
The Theorem 3.3 of \cite{Huet} that we employed above, does not work.
We should check the convergence of the critical pairs\\
(a) of the overlap between two reduction rules:
there is just one between \eqref{s2} and \eqref{s5},
and this converges, and\\
(b) between a reduction rule and an equality law,
there are some with an exchange law.\\
The critical pairs of (b) do not converge.
But if we enlarge the overlap term,
and take the two reducts of the enlarged overlap term,
then the two converge.
I do not know any theorem that would provide a simple proof of confluence from this.
(Perhaps a new challenge for term rewriters?)
As we cannot prove confluence,
we cannot establish equality $=$ for this program.
But we will prove an inequation with $\ree$.

\smallskip
Subtraction and addition obey some laws, as expected.
For all $x,y$ it should be something like
$y-y=\0$, $(x+y)-y=x$,
and $(x-y)+y=x$ for $\val{y}\leq\val{x}$,
where $\val{y}$ is the extensional value of $y$.
But these laws of subtraction/addition all contain two copies of $y$.
By ``locativity'' (see \ref{s:cond})
we have to distinguish them by naming them differently.
(And we can only prove $\ree$.)

\begin{prop}\label{p:xyy}
$\xyy \ree x$, for all \conus\ $x,y$.
\end{prop}
\proof
By induction on $x,y$.
There are five cases for the sum $x+y^0$.\\
We use another kind of exchange law:
\[ (A_0,A_1\ann)(B_0,B_1\ann)a \sme (A_0,B_1\ann)(B_0,A_1\ann)a \]
\textbf{(1)} $y=(Y\suc)y_0$:
\begin{alignat*}{2}
\xyy &\sme (x+(Y^0\suc)y_0^0)-(Y^1\suc)y_0^1, && \text{ by copy}\\
&\re (Y^0\suc)(x+y_0^0)-(Y^1\suc)y_0^1, && \text{ by \eqref{a1} and commut. of $+$}\\
&\re
 (Y^0,Y^1\ann)((x+y_0^0)-y_0^1), && \text{ by \eqref{s1}}\\
&\sme ((x+y_0^0)-y_0^1), && \text{ as $Y^0Y^{1-} = I$}\\
&\ree x, && \text{ by induction hyp.}
\end{alignat*}
\textbf{(2)} $y=(Y_0,Y_1\ann)y_0$:
\begin{alignat*}{2}
\xyy &\sme (x+(Y_0^0,Y_1^0\ann)y_0^0)-(Y_0^1,Y_1^1\ann)y_0^1, && \text{ by copy}\\
&\re (Y_0^0,Y_1^0\ann)(x+y_0^0)-(Y_0^1,Y_1^1\ann)y_0^1, && \text{ by \eqref{a2} and commut.\ of $+$}\\
&\re (Y_1^1,Y_0^1\ann)((Y_0^0,Y_1^0\ann)(x+y_0^0)-y_0^1) && \text{ by \eqref{s2}}\\
&\re (Y_1^1,Y_0^1\ann)(Y_0^0,Y_1^0\ann)((x+y_0^0)-y_0^1), && \text{ by \eqref{s5}}\\
&\sme (Y_1^1,Y_1^0\ann)(Y_0^0,Y_0^1\ann)((x+y_0^0)-y_0^1), && \text{ by the new exchange law}\\
&\sme (x+y_0^0)-y_0^1, && \text{ as $Y_0^0Y_0^{1-}=I$ and $Y_1^1Y_1^{0-}=I$}\\
&\ree x, && \text{ by induction hyp.}
\end{alignat*}
\textbf{(3)} $x=(X\suc)x_0$, $y=Y\0$:
\begin{alignat*}{2}
\xyy &\ree (X\suc)(x_0+Y^0\0)-Y^1\0, && \text{ by copy and \eqref{a3}}\\
&\re (X\suc)((x_0+Y^0\0)-Y^1\0), && \text{ by \eqref{s3}}\\
&\ree (X\suc)x_0 \ide x, && \text{ by induction hyp.}
\end{alignat*}
\textbf{(4)} $x=(X_0,X_1\ann)x_0$, $y=Y\0$: 
analogous to (3), use the rules \eqref{a4} and \eqref{s5}.

\medskip\noindent
\textbf{(5)} $x=X\0$, $y=Y\0$:
\begin{alignat*}{2}
\xyy &\sme (X\0+Y^0\0)-Y^1\0, && \text{ by copy}\\
&\re \brac{XY^0}\0-Y^1\0, && \text{ by \eqref{a5}}\\
&\sme \brac{XY^0}\0 - \brac{Y^1}\0, && \text{ as $Y^1\0\sme \brac{Y^1}\0$}\\
&\re \brac{\brac{XY^0}\brac{Y^1}^-}\0, && \text{ by \eqref{s4}}\\
&\sme \brac{\brac{XY^0}\brac{Y^{1-}}}\0 \\
&\sme \brac{\brac{XY^0Y^{1-}}}\0, && \text{ regard that $\limit\geq 3$}\\
&\sme \brac{\brac{X}}\0 \\
&\sme X\0  \ide x,  && \text{ regard that $\size(X)=1$}
\end{alignat*}
\qed

\section{System CN (constructed numbers): the arithmetic}\label{s:cn}

We have already given the algebra of conditions with the copy mechanism
and an explanation of the bracket mechanism in Subsection \ref{s:cond}.
Here we give the remaining rules on numbers,
we have seen the most important ones in applications in Section \ref{s:cnexp}.
We also give a detailed explanation of the atomic conditions.

We have already said that there seems to be great arbitrariness and freedom in choosing a 
way to give structure to Computation.
And following from this:
arbitrariness in choosing a definition of algorithm.
There is no a priori justification for the ``correctness'' of the choice.
The justification will come with the outcome of the approach (or not).
But it also seems that once the basic idea is fixed,
there is a prescribed way for working it out.
This way can only be seen with experience.
As I still have not enough experience with my own system,
it is here in a preliminary state,
it might still be ``incomplete''.

\medskip
We still have to explain:
What are the atomic conditions and where do they come from?
They have two possible sources:\\
\textbf{(1)}
There may be a main program-function on whose arguments the
whole Computation is ``grounded''.
Let these arguments correspond to the parameters $x,y,z$ of the function.
The atomic conditions are the conditions of the constructors
of the (arguments corresponding to the) $x,y,z$.
The $0$th constructor of $x$ is the $\0$, we name its condition by $x0$.
For $i\neq 0$, the $i$th constructor of $x$ is a $\suc$ or $\ann$.
For $\suc$ we name its condition by $xi$.
For $\ann$ we name its two conditions by $xi+$ and $xi-$.
Here is an example of an argument $x$ conditioned by its atomic conditions in this way:
\[ ((x3)\suc)((x2+),(x2-)\ann)((x1)\suc)((x0)\0) \]
\textbf{(2)}
It may be necessary to give atomic conditions for some of the constructors in the
right sides of the reduction rules for some function $f$.
We give them each a unique number $i$ and name them by $fi$.

The program for a function should be universally applicable,
so we use condition variables in its arguments,
and we see no atomic conditions of source (1).
They appear only if we want to ``ground'' the Computation,
which we will do in Section \ref{s:cnbasic}.

The complete algebra of conditions is in Subsection \ref{s:cond}.

\subsection*{The arithmetic of CN} $\phantom{x}$\\
The types $\ta$
 are: $\io$ (numbers), $\io^n$ ($n$-fold cartesian product),
$\io^n\fun \io^m$ (function space),
with $n,m\geq 1$ and $\io^1=\io$.

There are number variables $x,y,z$ for number (tuples), and function variables $f,g,h$.\\
There are typing environments $\Gamma$ and typing judgements
$\Gamma:a:\ta$ meaning
``under the typing environment $\Gamma$, $a$ has the type $\ta$''.
We mostly leave out the $\Gamma$ in these judgements,
as it is the same on both sides of a rule.

\smallskip
\noindent
Raw \textbf{Number terms} $a,b,c,d$:\\
$A,B$ are well-formed condition terms (\ie limited and with unique exponents)
with $\size(A)=\size(B)=1$.\\
$(\Gamma,x:\io^n):x:\io^n$\\
$A\0:\io$\\
$a:\io \rthen (A\suc)a : \io$\\
$a:\io \rthen (A,B\ann)a:\io$\\
$\aunio \rthen (\aun):\io^n$\\
$a:\io^n \rthen i\din a:\io$, for $i$ a natural number outside of the arithmetic, $1\leq i\leq n$\\
$a:\io^n \rthen \Acam: \io^n$\\
$a:\io^n \rthen a^0:\io^n \qquad
a:\io^n \rthen a^1:\io^n$\\
$f:\io^n\fun \io^m, \aunio \rthen f(\aun):\io^m$\\
Because of non-deterministic reduction,
there should be a sharing mechanism for number terms,
\ie constructs $a[v=b]$ where $a$ contains the special kind of variable $v$.
We leave this out.

We remember the Definitions \ref{d:cex} and \ref{d:ucex} of unique copy exponents.
\begin{defi}
A number term $a$ is \defin{well-formed}
if it has unique (copy) exponents 
and for every condition $A$ of a constructor in $a$ it is not $A=I$.
\end{defi}
[In the following every number term is supposed to be well-formed.]
\begin{defi}
Let $p$ be a position in term $t$.
The \defin{exponentiated subterm} of $t$ at position $p$ is
$t\csub p = (t/p)^{t\cex p}$.
\end{defi}
The following strong conjecture describes how the conditions of constructors in a number term are
distinguished:
\begin{conj}
Let $p \neq p'$ be positions of conditions of constructors in a (well-formed)
number term $a$.
Then not $a\csub p = a\csub p'$.
\end{conj}

\noindent
\textbf{Reduction rules for functions:}\\
Every used function variable $f:\io^n\fun \io^m$ has a 
finite set of associated \emph{reduction rules} of the form:
\[ f(\aun) \re b \text{ with $a_i:\io$ for all $i$, and $b:\io^m$} \]
The form of the $a_i$:\\
They are built from number variables $x:\io$ and constructors of the form
$A\0$, $(A\suc)$ or $(A,B\ann)$ 
where the $A,B$ are of the form $X$ or $\brac{X_1\dots X_j}$ with $j\geq 2$.
All variables appearing in $\aun$ are different (left-linearity).
(From this it already follows that the $a_i$ are well-formed.
Note also that $a_i$ does not contain atomic conditions.)\\
(The form $\brac{X_1\dots X_j}$ is useful for some programming tasks,
\eg for doubling a number.
It is a question if this form should be even more liberal.)\\
The form of $b$:\\
$b$ is a (well-formed) constructed number term.
Every (number or condition) variable in $b$ comes from the left side.
Every atomic condition in $b$ is of the form $fi$.
(Here $f$ is the function defined by the rule,
$i$ is unique for all the occurrences of atomic conditions in right sides of rules of $f$.)\\
For every $(A,B\ann)$ in $b$:
$A,B$ are  condition variables and $(A\suc), (B\suc)$
appear in the left side.
(It is not yet clear if the last restriction is ``needed''.)

\bigskip \noindent
We define the \textbf{smooth equality} $\sme$ on numbers.\\
There are rules that close $\sme$ to be a congruent equivalence.
\begin{align*}
A=B &\rthen A\0 \sme B\0 \\
A=B &\rthen (A\suc)a \sme (B\suc)a \\
A_0=B_0, A_1=B_1 &\rthen (A_0,A_1\ann)a \sme (B_0,B_1\ann)a \\
A=B &\rthen \Acam \sme B\co a
\end{align*}
\begin{align*}
(A\suc)(B\suc)a &\sme (B\suc)(A\suc)a \\
(A_0,A_1\ann)(B\suc)a &\sme (B\suc)(A_0,A_1\ann)a \\
(A_0,A_1\ann)(B_0,B_1\ann)a &\sme (B_0,B_1\ann)(A_0,A_1\ann)a \\
(A_0,A_1\ann)(B_0,B_1\ann)a &\sme (A_0,B_1\ann)(B_0,A_1\ann)a \\
(A\suc)(B_0,B_1\ann)a &\sme (B_0\suc)(A,B_1\ann)a
\end{align*}
\begin{align*}
(A\0)^0 &\sme A^0\0 &
(A\0)^1 &\sme A^1\0 \\
((A\suc)a)^0 &\sme (A^0\suc)a^0 &
((A\suc)a)^1 &\sme (A^1\suc)a^1 \\
((A,B\ann)a)^0 &\sme (A^0,B^0\ann)a^0 & 
((A,B\ann)a)^1 &\sme (A^1,B^1\ann)a^1 \\
(\aun)^0 &\sme (a_1^0,\dots, a_n^0) & 
(\aun)^1 &\sme (a_1^1,\dots, a_n^1) \text{ for } n\geq 2 
\end{align*} 
Here the four equations on numbers before Proposition \ref{p:heck} should be inserted.

\medskip\noindent
The following equations will later have a special status
because of their asymmetric character:\\
\textbf{Tuple-selection:}\\
$i\din (\aun) \sme a_i$, for $1\leq i\leq n$\\
\textbf{Copy-expansion:}
\begin{align*}
A\co (B\0) &\sme \brac{AB}\0 \\
A\co (B\suc)a &\sme (\brac{A^0B}\suc)(A^1\co a) \\
A\co (B,C\ann)a &\sme (\brac{A^{00}B},\brac{A^{01}C}\ann)(A^1\co a)
\end{align*}
\textbf{Inversion-simplification:}\\
$(A,B\ann)a \sme a$, for $AB^- = I$

\bigskip\noindent
We define the \textbf{equality reduction} $\ree$ on numbers.\\
There are rules that close $\ree$ to be reflexive, transitive, congruence.\\
Let $f(\aun)\re b$ be a reduction rule for the function $f$.\\
Let $\sigma$ be a substitution of the variables of $(\aun)$
by number \resp condition terms
such that for every condition variable $X$ it is
$\size(\sigma(X))=1$.\\
Then $\sigma(f(\aun))\ree \sigma(b)$,
where the substitution extends to term arguments.\\
$\sigma(f(\aun))$ must be well-formed.\\
$\ree$ contains $\sme$: 
$a\sme b \rthen a\ree b$.

\bigskip\noindent
We define the \textbf{direct equality reduction} $\wree$ on numbers.\\
$\wree$ is built up like $\ree$, we describe this roughly.
Take all the defining equations and reductions of $\ree$,
but with the following exemptions:\\
$A^0A^{1-}=I$, $A^1A^{0-}=I$ on conditions are not allowed.\\
$A=A^0A^1$ on conditions is only allowed from left to right.\\
The equations of tuple-selection and copy-expansion (for $\sme$) are only allowed from left to right.\\
The equation of inversion-simplification (for $\sme$) is not allowed.\\
Then close $\dree$ to be reflexive, transitive, congruence.

Equality reduction $\ree$ accounts for the full possibilities of Computation.
But direct equality reduction restricts the Computation to steps that have
symmetric character or that go ``straight forward'',
so that Computation makes no ``detours''.

\begin{prop}
Let $a$ be a well-formed number term and
$a\ree b$ for some $b$.
Then $b$ is also well-formed.
\end{prop}
\proof
Check the defining equations of $\ree$.
Also the reduction rules $L\re R$ preserve well-formedness (unique copy exponents),
as $R$ is well-formed.
\qed

\section{Basic notions of the theory of system CN}\label{s:cnbasic}

Because of lack of time, I cannot develop the theory of CN properly.
We give here only some basic definitions,
namely of CN-algorithm and of direct algorithm.

\begin{defi}
A \defin{constructor number} is a number term $a:\io$ built entirely of
constructors $A\0, B\suc, (A,B\ann)$,
without any condition variables in the conditions.
(So the conditions are built only of atomic conditions and condition operators.)\\
$\Con$ is the set of all constructor numbers.\\
$\Conc$ is the set of all equivalence classes $\smec{a}$ \wrt $\sme$,
for $a\in \Con$.\\
$\Conc^n$ is the $n$-fold cartesian product,
it is isomorphic to the set of all equivalence classes $\smec{a}$ for $a\in \Con^n$.\\
A \emph{ground number} for a number variable $x:\io$ is a constructor number $a$
that is formed as the example described in the explanation of atomic conditions
at the beginning of Section \ref{s:cn}, point (1).\\
For $n\geq 1$, $\Gr^n$ is the set of all $(\aun)$ with $a_i$
a ground number for the (fixed) variable $x_i$.
\end{defi}

\begin{defi}\label{d:algo}
Let $f:\io^n\fun \io^m$ be a function (symbol) defined by a program.\\
The \defin{algorithm of} $f$ is the map $\algor(f): \Gr^n\fun \mathit{power}(\Conc^m)$ defined by
\[ \algor(f)a = \set{\smec{b}}{b\in \Con^m \text{ and } fa\ree b}.\]
We call such algorithm of $f$ a CN-algorithm.\\
For CN-algorithms $F,G: \Gr^n \fun \mathit{power}(\Conc^m)$ we define the partial order\\
$F\subal G$ if for all $a\in \Gr^n$: $Fa\sub Ga$.
\end{defi}

Note that the arguments of program-functions and algorithms, as we define them,
are complete numbers, not partial numbers in the sense of denotational semantics.\\
We have the strong conjecture that the functions
$f(x,y)\re \xyy$ and $g(x,y)\re x$,
with $+,-$ of Section \ref{s:cnexp},
describe the same CN-algorithm.
This would be proved by Proposition \ref{p:xyy}
if we had a proof of confluence of the reductions for $+,-$.
But also with the non-confluent variant it should be possible to prove it.\\
(If our definitions would admit also partial numbers as arguments,
then surely $f$ and $g$ would be different algorithms,
as $f$ is less defined than $g$.)

\bigskip\noindent
It will surely be possible to give for CN a categorical structure with
the natural numbers as objects and the algorithms
$F:\Gr^n\fun \mathit{power}(\Conc^m)$ as morphisms from $n$ to $m$.
What would such a category look like in the general case 
that encompasses also other data structures?
Do such categories admit transformation laws like those
of Noson Yanofsky's category \cite{Yanofsky}?

With the direct equality reduction $\dree$ we have the possibility to specify
when a program or CN-algorithm does only straight forward (direct) work
and does not reverse a step that was once taken.

\begin{defi}\label{d:direct}
Let $f:\io^n\fun\io^m$ be a function(symbol) defined by a program.\\
$f$ is \defin{direct} if for all $a\in \Gr^n$, $b\in \Con^m$ with $fa\ree b$
there is some $b' \in \Con^m$ with $fa\dree b'$ and $b\sme b'$.\\
A CN-algorithm is \defin{direct} if it is the algorithm of a direct $f$.
\end{defi}

All these notions should have corresponding notions in the realm of normal programs.
This demands a notion of ``correctness'' of a CN-program with respect to a normal program.
This is not yet solved.

\section{Outlook}

We list some questions that are left open:\\
Is there a meaningful notion of correctness of a CN-program \wrt a normal program?
This would enable us to translate the notion of CN-algorithm to the realm of normal programs.\\
More examples are needed, \esp examples that use the bracket mechanism in an
essential way to prove equivalence of programs.\\
How does the system look like for other data structures than tally numbers?\\
What is the categorical structure of the system for such general data structures?\\
Can we make the system reversible?
We have imposed the restriction of unique copying.
What about the restriction that no condition gets lost?\\
What are the applications of the notion of direct Computation,
given in Definition  \ref{d:direct}?\\
Are there intensional versions of Hoare program logic and Scott denotational semantics
for system CN?\\
Can we ground a ``more qualitative'', not merely quantitative complexity theory on our system?
This could lead to new complexity measures.\\
More speculative:
What are the wider possibilities of encoding properties in conditions,
\esp when we see the propositional conditions in the concept objects,
speculated upon in Section 3?
Can we encode in an algorithm a complexity property of itself,
compare the G\"odel sentence?

\section*{Acknowledgement}
I thank Reinhold Heckmann for very carefully reading the drafts of this paper and many fruitful discussions
and hints. He prevented some severe formal mistakes. \\
I thank Tobias M\"omke and Reinhard Wilhelm for their interest in this work
and suggestions to improve the paper.\\
I thank Sebastian Hack and the members of his chair for their hospitality and support,
\esp Roland Lei\ss a, Klaas Boesche and Michael Jacobs for help with the computer.
\bibliography{bib}

\begin{thebibliography}{10}

\bibitem{Bergstra/Middelburg}
J.A. Bergstra and C.A. Middelburg.
\newblock On algorithmic equivalence of instruction sequences for computing bit
  string functions.
\newblock arXiv:1402.4950v3.

\bibitem{Cousot}
Patrick Cousot and Radhia Cousot.
\newblock Abstract interpretation: A unified lattice model for static analysis
  of programs by construction or approximation of fixpoints.
\newblock In {\em 4th POPL}, pages 238--252, 1977.

\bibitem{Gasarch:1}
William~I. Gasarch.
\newblock The {P}=?{NP} poll.
\newblock {\em SIGACT News}, 33:34--47, 2002.

\bibitem{Gasarch:2}
William~I. Gasarch.
\newblock The second {P}=?{NP} poll.
\newblock {\em SIGACT News}, 43:53--77, 2012.

\bibitem{Girard:goii}
Jean-Yves Girard.
\newblock Geometry of interaction: Interpretation of system {F}.
\newblock In J.W. Gray and A.~Scedrov, editors, {\em Proc. Logic Colloquium
  88}, pages 221--260. North Holland, 1989.

\bibitem{Girard:tgoi}
Jean-Yves Girard.
\newblock Towards a geometry of interaction.
\newblock In J.W. Gray and A.~Scedrov, editors, {\em Categories in Computer
  Science and Logic}, pages 69--108, 1989.

\bibitem{Girard:locus}
Jean-Yves Girard.
\newblock Locus {S}olum. {F}rom the rules of logic to the logic of rules.
\newblock {\em Mathematical Structures in Computer Science}, 11:301--506, 2001.

\bibitem{Girard:locat}
Jean-Yves Girard.
\newblock Normativit\'e, locativit\'e et identit\'e.
\newblock \url{iml.univ-mrs.fr/~girard/prelude}, 2009.

\bibitem{Haghverdi/Scott}
Esfandiar Haghverdi and Philip Scott.
\newblock Geometry of interaction and the dynamics of proof reduction: a
  tutorial.
\newblock \url{www.site.uottawa.ca/~phil/papers/HS.GoI-tut.33.pdf}, 2008.

\bibitem{Hartmanis/Stearns}
Juris Hartmanis and Richard~E. Stearns.
\newblock On the computational complexity of algorithms.
\newblock {\em Transactions of the American Mathematical Society},
  117:285--306, 1965.

\bibitem{Hoare}
C.A.R. Hoare.
\newblock An axiomatic basis for computer programming.
\newblock {\em Comm. ACM}, 12:576--580, 1969.
\newblock reprinted in C.A.R. Hoare: Essays in computing science, Prentice Hall
  1989.

\bibitem{Huet}
G.~Huet.
\newblock Confluent reductions: Abstract properties and applications to term
  rewriting systems.
\newblock {\em J. of the ACM}, 27(4):797--821, 1980.

\bibitem{Moschovakis:found}
Yiannis~N. Moschovakis.
\newblock On founding the theory of algorithms.
\newblock In H.G. Dales and G.~Oliveri, editors, {\em Truth in Mathematics},
  pages 71--104. Clarendon Press, 1998.

\bibitem{Moschovakis:what}
Yiannis~N. Moschovakis.
\newblock What is an algorithm?
\newblock In {\em Mathematics Unlimited -- 2001 and beyond}. Springer, 2001.

\bibitem{Scott}
Dana~S. Scott.
\newblock A type-theoretical alternative to {ISWIM}, {CUCH}, {OWHY}.
\newblock {\em Theoretical Computer Science}, 121:411--440, 1993.
\newblock Originally written and distributed in 1969.

\bibitem{Turing}
Alan~M. Turing.
\newblock On computable numbers, with an application to the
  {E}ntscheidungsproblem.
\newblock {\em Proceedings of the London Mathematical Society}, 42,series
  2:230--265, 1936--1937.

\bibitem{Oostrom/Vrijer}
Vincent van Oostrom and Roel de~Vrijer.
\newblock Equivalence of reductions.
\newblock In Terese, editor, {\em Term Rewriting Systems}, pages 301--474.
  Cambridge University Press, 2003.

\bibitem{Yanofsky}
Noson~S. Yanofsky.
\newblock Towards a definition of an algorithm.
\newblock {\em Journal of Logic and Computation}, 2010.
\newblock arXiv:0602053v3.

\end{thebibliography}
\bibliographystyle{plain}
\end{document}